\documentclass[lettersize,journal]{IEEEtran}
\usepackage{amsmath,amssymb,amsfonts}
\usepackage{caption}
\usepackage{algorithmic}
\usepackage[ruled,linesnumbered]{algorithm2e}
\usepackage{array}

\usepackage{textcomp}
\usepackage{stfloats}
\usepackage{subcaption}
\usepackage{url}
\usepackage{verbatim}
\usepackage{graphicx}
\usepackage{textcomp}
\usepackage{xcolor}
\usepackage{cite}
\usepackage{hyperref} 
\hypersetup{
    colorlinks=true, 
    citecolor=blue,
    linkcolor=blue,   
    filecolor=blue,     
    urlcolor=black       
}
\begin{document}

\title{Cognitive Digital Twins for Self-Aware Channel Estimation}

\author{Afan Ali, Ali Arshad Nasir,~\IEEEmembership{Senior Member,~IEEE}, and Daniel Benevides da Costa,~\IEEEmembership{Senior Member,~IEEE}
\thanks{The authors are with the Interdisciplinary Research Center for Communication Systems and Sensing (IRC-CSS), Department of Electrical Engineering, King Fahd University of Petroleum and Minerals (KFUPM), Dhahran 31261, Saudi Arabia (e-mail: afan.ali@kfupm.edu.sa; anasir@kfupm.edu.sa; danielbcosta@ieee.org).}}
\vspace{-5pt}

% The paper headers
\markboth{Journal of \LaTeX\ Class Files,~Vol.~14, No.~8, July~2026}%
{Shell \MakeLowercase{\textit{et al.}}: A Sample Article Using IEEEtran.cls for IEEE Journals}

\maketitle

\begin{abstract}
Artificial intelligence (AI) and machine learning (ML)-based channel estimators silently degrade when propagation conditions drift from their training distributions. This letter proposes a model-agnostic cognitive digital twin (CDT) framework that combines a variational autoencoder (VAE) with latent activation monitoring to detect distribution drift and autonomously execute \textsc{continue}, \textsc{update}, or \textsc{retire} lifecycle actions without requiring ground-truth channel knowledge. The proposed framework is fully compatible with the AI-native lifecycle management envisioned in 3rd Generation Partnership Project (3GPP). Simulations over various channels demonstrate accurate drift detection and robust channel estimation, consistently outperforming conventional offline-trained deep learning estimators under moderate and severe channel drift. 
\end{abstract}


\section{Introduction}
\label{sec:intro}

\IEEEPARstart{T}{he} transition toward Artificial Intelligence (AI)-native
air interfaces in 5G-Advanced and 6G has made learned channel estimators
viable for practical deployment. In 3rd Generation Partnership Project
(3GPP) Release~19, channel estimation enhancement and AI model lifecycle
management have both emerged as active standardization topics, reflecting
the need to ensure reliable operation post-deployment, not just improved
accuracy~\cite{3gpp_tr38843}. Recent deep learning approaches, including
residual, attention-based, and transformer
architectures~\cite{attrnet,sccenet,channelformer}, consistently outperform
classical least squares (LS) and linear minimum mean square error (LMMSE)
estimators, but implicitly assume the deployment channel matches the
training distribution.

In practice, wireless propagation continuously evolves as users move across
environments, such as, a model trained under Extended Pedestrian (EPA) conditions may later operate under Extended Vehicular A (EVA) or Extended Typical Urban
(ETU) statistics, but still continue producing estimates without any indication
that its reliability has deteriorated~\cite{attrnet,sccenet}. This silent degradation is evident
across recent estimators, for example, attention mechanism and residual
network (AttRNet) and symmetric convolutional neural networks (CNN) based channel estimation network (SCCENet) target accuracy alone
\cite{attrnet,sccenet}, recurrent estimators degrade once channel dynamics exceed their design assumptions \cite{gru_gizzini2024,keykhosravi2026}, and Channelformer flags autonomous online adaptation as an open challenge \cite{channelformer}. None of these techniques provide a mechanism to detect distribution shift or determine when adaptation is needed.

Digital twin (DT) technology offers a natural foundation for this problem by continuously linking the physical channel with its virtual
counterpart~\cite{njoku2025battery,bae2025dt}, and recent work on Generative
Digital Twin Channels shows that generative AI (GAI) can model wireless channel
distributions and synthesize realizations for new environments~\cite{gdtc_liu2025}. Variational autoencoders (VAEs) are further established as anomaly
detectors in DT systems~\cite{chen2024gai_hdt}, yet existing wireless VAE
applications target communication or sensing tasks rather than deployment
reliability~\cite{nemati2023vqvae,adhikary2024hmimo,yang2026genai}. In particular, the
closest related work in~\cite{clce_kong2025} detects distribution changes via a contrastive VAE, but requires environment-labelled data, maintains an unbounded model ensemble, and lacks structured lifecycle decisions.
Consequently, no existing channel estimator jointly achieves label-free drift detection, 3GPP-aligned lifecycle management, and autonomous model replacement.

In this letter, we propose a model-agnostic cognitive digital twin (CDT) framework that equips AI and machine learning (ML) channel estimators with continuous self-monitoring and
autonomous lifecycle management. While lifecycle management is a standardized concept in 3GPP~\cite{3gpp_tr38843}, no standardized mechanism exists to trigger it autonomously. The proposed CDT provides this mechanism by
combining two complementary monitoring signals, i.e., a VAE operating
directly on received pilots to detect input distribution shift and an
activation drift detector that monitors changes in internal neural
representations. Their outputs are fused by a lifecycle controller that
autonomously selects the \textsc{continue}, \textsc{update}, or
\textsc{retire} actions defined in 3GPP TR~38.843~\cite{3gpp_tr38843}. When
retirement is required, a vector quantization (VQ)-VAE-based generative
pipeline characterizes the new propagation environment, synthesizes
additional training samples, and validates a replacement estimator. The main contributions of this letter can be summarized as follows:

\begin{itemize}
\item It is proposed a CDT framework that couples AI/ML channel estimation with autonomous lifecycle management to detect and respond to channel distribution drift, addressing the silent performance degradation commonly observed in deployed estimators.

\item To the best of our knowledge, this is the first work to employ VAE reconstruction error on received pilot observations as a ground-truth-free drift indicator. Combined with latent activation monitoring, it enables robust detection of both input-space and representation-space distribution shifts.

\item It is developed a GAI-driven VQ-VAE pipeline for autonomous model replacement using synthetic data generated from limited post-drift observations, a capability not available in prior drift-aware channel estimation.
\end{itemize}

\section{System Model and Problem Statement}
\label{sec:system}

\subsection{Signal Model}
\begin{figure}[t]
\centering
\includegraphics[width=\columnwidth]{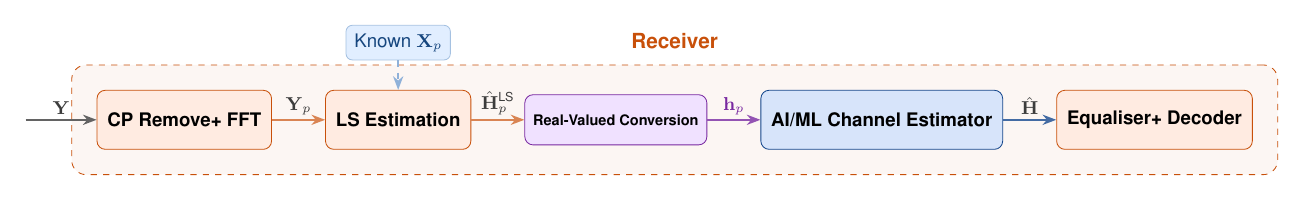}
\caption{Receiver system model.}
\label{fig:system_model}
\end{figure}

We consider a single-input single-output (SISO) orthogonal frequency-division multiplexing (OFDM) system with $N_f$ subcarriers and $N_s$ OFDM symbols per slot~\cite{sccenet}. After cyclic prefix (CP) removal and Discrete Fourier transform (DFT) processing, the received signal on subcarrier $k$ and OFDM symbol $\ell$ can be formulatd as $Y[k,\ell]=H[k,\ell]X[k,\ell]+W[k,\ell]$, where $H[k,\ell]$ denotes the channel frequency response of the $\ell$-th symbol on $k$-th subcarrier, $X[k,\ell]$ is the transmitted data symbol, and $W[k,\ell]\sim\mathcal{CN}(0,\sigma_n^2)$ is additive white Gaussian noise. Stacking of all subcarriers and symbols can be written as
\begin{equation}
\mathbf{Y} = \mathbf{H}\circ\mathbf{X} + \mathbf{W},
\label{eq:signal_stack}
\end{equation}
where $\circ$ denotes the Hadamard product. The complete receiver pipeline, including LS estimation, preprocessing, and AI/ML-based channel estimation, is illustrated in Fig.~\ref{fig:system_model}.

\subsection{Pilot Structure and LS Estimation}

Pilots are inserted following a comb-type demodulation reference signal (DMRS) pattern over $N_{ps}$ pilot OFDM symbols with pilot spacing $P_s$, resulting in $N_{pf}$ pilot subcarriers per pilot symbol and $N_p=N_{pf}N_{ps}$ pilot resource elements per slot. Collecting the received signal, transmitted pilots, channel, and noise at all pilot positions into $\mathbf{Y}_p,\mathbf{X}_p,\mathbf{H}_p,\mathbf{W}_p\in\mathbb{C}^{N_{pf}\times N_{ps}}$, the pilot observation model can be expressed as
\begin{equation}
\mathbf{Y}_p = \mathbf{H}_p\circ\mathbf{X}_p + \mathbf{W}_p.
\label{eq:pilot_model}
\end{equation}
Since $\mathbf{X}_p$ is known at the receiver, the LS estimate can be obtained as
\begin{equation}
\hat{\mathbf{H}}_p^{\rm LS}
=
\mathbf{Y}_p \oslash \mathbf{X}_p,
\label{eq:ls_estimate}
\end{equation}
where $\oslash$ denotes element-wise division. The resulting estimate is vectorized by stacking its real and imaginary components to form the input of the AI/ML estimator.

\subsection{Problem Statement}
An AI/ML estimator $f_{\boldsymbol{\theta}}:\mathbb{R}^{2N_p}\to\mathbb{R}^{2N_fN_s}$, which is realized as a neural network, such as, a convolutional or fully-connected
architecture and maps the LS pilot estimate
$\mathbf{h}_p$ to the full channel, is trained offline on channel realizations drawn from a training distribution $p_{\rm train}(\mathbf{H})$ by minimizing
\begin{equation}
\mathcal{L}_{\rm train}(\boldsymbol{\theta}) = \mathbb{E}_{\mathbf{H}\sim p_{\rm train}} \left\| f_{\boldsymbol{\theta}}(\mathbf{h}_p) - {\rm vec}_{\mathbb{R}}(\mathbf{H}) \right\|_2^2,
\label{eq:train_loss}
\end{equation}
where ${\rm vec}_{\mathbb{R}}(\mathbf{H})\in\mathbb{R}^{2N_fN_s}$ is  real-valued vectorization of the full channel matrix. The benchmark normalized mean square error (NMSE) under matched conditions can be written as
\begin{equation}
{\rm NMSE}^* = \frac{\mathbb{E}_{p_{\rm train}}\left\|\hat{\mathbf{H}}-\mathbf{H}\right\|_F^2}{\mathbb{E}_{p_{\rm train}}\left\|\mathbf{H}\right\|_F^2}.
\label{eq:nmse_star}
\end{equation}
During deployment, suppose the channel distribution shifts to $p_{\rm deploy}(\mathbf{H})\neq p_{\rm train}(\mathbf{H})$. The actual NMSE at time slot $t$ can be expressed as
\begin{equation}
{\rm NMSE}(t) = \frac{\left\| f_{\boldsymbol{\theta}}\left(\mathbf{h}_p^{(t)}\right) - {\rm vec}_{\mathbb{R}}\left(\mathbf{H}^{(t)}\right)\right\|_2^2}{\left\|{\rm vec}_{\mathbb{R}}\left(\mathbf{H}^{(t)}\right)\right\|_2^2}.
\label{eq:nmse_t}
\end{equation}
Since $\mathbf{H}^{(t)}$ is unavailable at the receiver during inference, ${\rm NMSE}(t)$ cannot be evaluated and the estimator has no direct means to assess whether its output is reliable. This motivates the following definition.

\textbf{Definition 1 (Silent Degradation).} An estimator $f_{\boldsymbol{\theta}}$ suffers silent degradation at slot $t$ if ${\rm NMSE}(t)\gg{\rm NMSE}^*$ while no observable signal at the receiver provides any indication of this deterioration, i.e., the estimator continues producing outputs $\hat{\mathbf{H}}^{(t)}$ without any internal reliability flag. Since $\mathbf{H}^{(t)}$ grows increasingly dissimilar from the training mean under distribution shift, ${\rm NMSE}(t)$ increases correspondingly; however this deviation remains unobservable since $\mathbf{H}^{(t)}$ is unknown, motivating the search for an observable proxy.

\begin{figure}[t]
\centering
\includegraphics[width=\columnwidth]{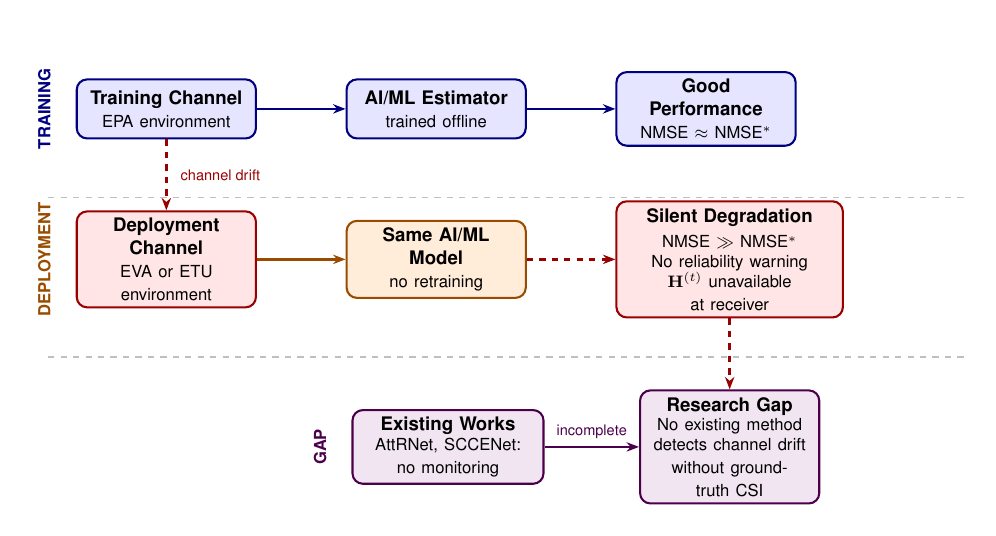}
\caption{Silent degradation problem in the existing literature.}
\label{fig:problem}
\end{figure}

\textbf{Observation 1.} A distributional shift in $\mathbf{H}^{(t)}$ induces a corresponding shift in the marginal distribution of the received pilot vector $\mathbf{y}_p^{(t)}\in\mathbb{R}^{2N_p}$. Since $\mathbf{y}_p^{(t)}=\mathbf{h}_p^{(t)}+\mathbf{w}_p^{(t)}$ is a deterministic function of $\mathbf{H}_p^{(t)}$ corrupted by noise, if $p_{\rm deploy}(\mathbf{H})\neq p_{\rm train}(\mathbf{H})$, it follows that
\begin{equation}
D_{\rm KL}\big(p_{\rm deploy}(\mathbf{y}_p)\,\big\|\,p_{\rm train}(\mathbf{y}_p)\big) > 0,
\label{eq:kl_shift}
\end{equation}
where $D_{\rm KL}(\cdot\|\cdot)$ denotes the Kullback--Leibler (KL) divergence, which measures the statistical discrepancy between the deployment and training pilot distributions. Rather than evaluating ${\rm NMSE}(t)$ directly, the proposed system, therefore, monitors the statistical behaviour of the observable pilot vector $\mathbf{y}_p^{(t)}$ as a proxy for distribution shift; a persistent increase in the divergence of \eqref{eq:kl_shift} signals that the estimator is operating outside its training distribution and motivates a lifecycle action. Fig.~\ref{fig:problem} summarizes the silent degradation problem and research gap. 

\section{Proposed Cognitive Digital Twin Framework}
\label{sec:framework}

In this section, we present the proposed CDT framework. Drawing on the three-space DT architecture of~\cite{njoku2025battery,bae2025dt}, the CDT organizes the AI/ML estimator, self-monitoring module, and lifecycle controller within a unified structure.

\subsection{CDT Architecture Overview}

\begin{figure}[t!]
     \centering
     \begin{subfigure}[b]{0.35\textwidth}
         \centering
         \includegraphics[width=\textwidth]{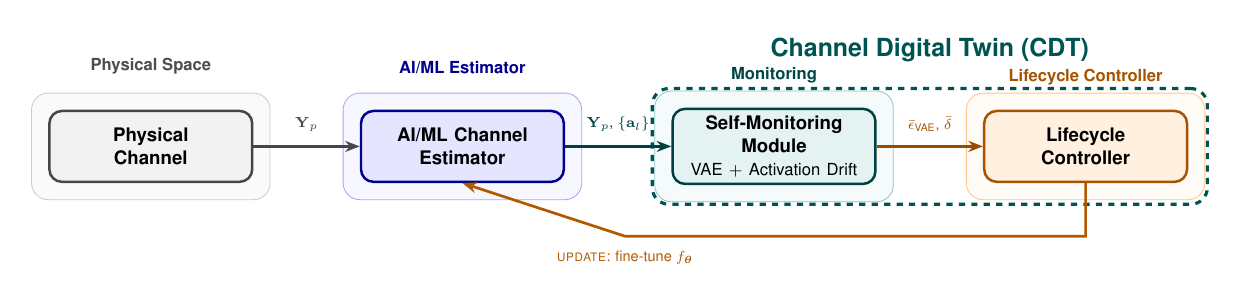}
         \caption{Overall system architecture.}
         \label{fig:3a}
     \end{subfigure}
     \hfill
     \begin{subfigure}[b]{0.35\textwidth}
         \centering
         \includegraphics[width=\textwidth]{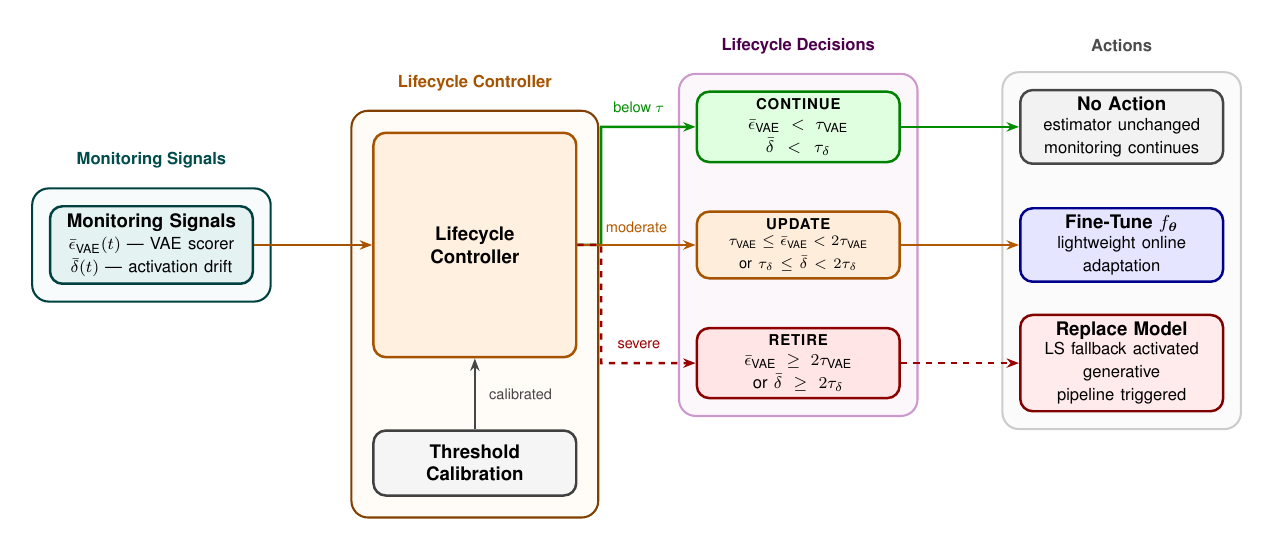}
         \caption{Illustration of Lifecycle controller.}
         \label{fig:3b}
     \end{subfigure}
        \caption{Proposed framework.}
        \label{fig_framework}
\end{figure}

As illustrated in Fig.~\ref{fig:3a}, the CDT comprises a physical space
housing the real wireless channel and a digital space containing the AI/ML estimator, self-monitoring module, and lifecycle controller. The
self-monitoring module fuses VAE-based input scoring and activation drift detection into smoothed signals, which the controller maps to structured actions, such as, moderate drift triggers fine-tuning of $f_{\boldsymbol{\theta}}$,
while severe drift activates the generative pipeline to deploy a replacement $f_{\boldsymbol{\theta}'}$. Decision logic, thresholds, and resulting actions are detailed in Fig.~\ref{fig:3b}..

\subsection{AI/ML Channel Estimator}

The proposed framework is model-agnostic which can be integrated with both lightweight and high-capacity AI/ML channel estimators. In this work, a four-layer fully convolutional network (FCN) is adopted as the default backbone, since this depth was found sufficient to capture the LS-to-channel mapping with diminishing NMSE gains beyond four layers, while keeping the per-layer activation dimensionality low enough to make activation-based drift
monitoring computationally lightweight. More powerful architectures, such as AttRNet~\cite{attrnet} and SCCENet~\cite{sccenet}, can also be incorporated without modification, provided they expose intermediate layer activations, yielding improved estimation performance at the expense of increased computational complexity. The resulting activations, $\{\mathbf{a}_l\}_{l=1}^{L}$, are forwarded to the VAE-based monitoring module during inference with negligible additional overhead.

\subsection{Self-Monitoring Module}
\label{subsec:monitoring}
The self-monitoring module produces two 
complementary reliability indicators from 
receiver-observable quantities without 
ground-truth channel, motivated by the established role of VAEs as anomaly 
detectors in DT 
architectures~\cite{chen2024gai_hdt}.
\subsubsection{VAE-Based Input Distribution Scoring}
A VAE $\mathcal{V}_{\boldsymbol{\phi}}$ can be trained 
on received pilot observations $\mathbf{y}_p \in 
\mathbb{R}^{2N_p}$ collected under 
training-matched conditions by minimizing the 
evidence lower bound as follows
\begin{equation}
\mathcal{L}_{\text{VAE}} = 
\underbrace{\mathbb{E}\left[\left\|\mathbf{y}_p - 
\hat{\mathbf{y}}_p\right\|^2\right]}_{\text{reconstruction}} 
+ \beta \cdot \underbrace{D_{\text{KL}}\!\left(
\mathcal{N}(\boldsymbol{\mu},\boldsymbol{\sigma}^2)
\,\|\,\mathcal{N}(\mathbf{0},\mathbf{I})\right)}_{
\text{regularization}},
\label{eq:vae_loss}
\end{equation}
where $\beta$ balances reconstruction fidelity 
against latent regularity. Rather than estimating 
$p_{\rm deploy}(\mathbf{y}_p)$ directly, which would 
need prohibitively many samples in $2N_p$ dimensions, 
$\mathcal{V}_{\boldsymbol{\phi}}$ is pre-trained once on 
$p_{\rm train}(\mathbf{y}_p)$ and scores each new pilot 
observation individually against it, so drift is flagged 
from single samples rather than a batch-estimated 
distribution. The reconstruction error can be written as 
\begin{equation}
\epsilon_{\text{VAE}}(t) = \left\|\mathbf{y}_p^{(t)} 
- \hat{\mathbf{y}}_p^{(t)}\right\|^2,
\label{eq:vae_error}
\end{equation}
which is exponentially weighted moving average~(EWMA)-smoothed over a sliding window (Section~\ref{subsec:drift_detection}) to give a robust drift 
indicator from a modest observation stream rather than a large pre-collected sample set. It serves as a familiarity 
score during deployment, where low values indicate 
distributional consistency and elevated values signal departure. Moreover, VAE operates on raw $\mathbf{y}_p$ rather 
than on pilots reconstructed from $\hat{\mathbf{H}}$, 
since the latter would simply recover the LS 
residual already minimized in 
Section~\ref{sec:system}, yielding no useful 
drift information. This non-circular design is 
absent from all prior wireless VAE 
applications~\cite{nemati2023vqvae,
adhikary2024hmimo,yang2026genai}.

\subsubsection{Activation Statistics Drift Detection}
\label{subsec:drift_detection}

The second signal detects changes in the 
estimator's internal behaviour motivated by the 
empirical finding of~\cite{channelformer} that 
network representations shift measurably across 
EPA, EVA, and ETU conditions. Reference 
statistics can be recorded per layer $l$ during 
initial deployment as
\begin{equation}
\boldsymbol{\mu}_l^{\text{ref}} = 
\mathbb{E}\left[\mathbf{a}_l\right], \quad 
\boldsymbol{\sigma}_l^{\text{ref}} = 
\text{std}\left[\mathbf{a}_l\right],
\label{eq:ref_stats}
\end{equation}
and the drift score over a sliding window of 
$W$ observations can be expressed as 
\begin{equation}
\delta_l(t) = \frac{1}{d_l}\sum_{j=1}^{d_l} 
\frac{\left|\mu_{l,j}^{(t)} - 
\mu_{l,j}^{\text{ref}}\right|}
{\sigma_{l,j}^{\text{ref}} + \epsilon},
\label{eq:drift_score}
\end{equation}
where $d_l$ is the layer dimension and 
$\epsilon = 10^{-8}$ is a small constant 
that prevents division by zero when a 
neuron's reference standard deviation 
$\sigma_{l,j}^{\text{ref}}$ is negligibly 
small due to near-constant activation 
across the calibration window. The 
overall score $\delta(t) = \max_{l}\,\delta_l(t)$ 
takes the layer-wise maximum since earlier layers 
respond to amplitude changes while deeper layers 
respond to correlation structure changes, and 
averaging would suppress localised shifts. The 
two signals are complementary in that the VAE 
responds rapidly to broad input transitions 
while activation drift detects subtler internal 
changes.

\subsection{Autonomous Lifecycle Controller}
\label{subsec:lifecycle}
The lifecycle controller applies EWMA
smoothing with factor $\alpha \in (0,1)$ to the 
$\epsilon_{\text{VAE}}(t)$ and  $\delta(t)$ defined in
\eqref{eq:vae_error} and \eqref{eq:drift_score}, giving the smoothed signals
\begin{align}
\bar{\epsilon}_{\text{VAE}}(t) &= \alpha\, 
\epsilon_{\text{VAE}}(t) + 
(1-\alpha)\,\bar{\epsilon}_{\text{VAE}}(t-1),
\label{eq:ewma_vae}\\
\bar{\delta}(t) &= \alpha\,\delta(t) + 
(1-\alpha)\,\bar{\delta}(t-1).
\label{eq:ewma_delta}
\end{align}
These signals are compared against calibrated thresholds $\tau_{\text{VAE}}$ and
$\tau_{\delta}$,
\begin{equation}
\tau_{\text{VAE}} = \bar{\epsilon}_{\text{VAE}}^{\,
\text{cal}} + 2\,\sigma_{\epsilon}^{\text{cal}},
\qquad
\tau_{\delta} = \bar{\delta}^{\,\text{cal}} + 
2\,\sigma_{\delta}^{\text{cal}},
\label{eq:thresholds}
\end{equation}
where $\bar{\epsilon}_{\text{VAE}}^{\,\text{cal}}$, 
$\bar{\delta}^{\,\text{cal}}$ and 
$\sigma_{\epsilon}^{\text{cal}}$, 
$\sigma_{\delta}^{\text{cal}}$ denote the sample 
means and standard deviations of each signal, 
collected over a calibration period of $T_{\text{cal}}$ slots under 
matched conditions, setting each threshold two standard deviations 
above the mean yields an empirical false-alarm rate of approximately 
$2.3\%$ without manual tuning. This factor of two further defines, for 
each threshold $\tau \in \{\tau_{\text{VAE}}, \tau_\delta\}$, a moderate drift 
region $[\tau,\,2\tau)$ triggering \textsc{update} and a severe region 
$[2\tau,\,\infty)$ triggering \textsc{retire}, separating recoverable 
shift from irrecoverable mismatch. As shown in Fig.~\ref{fig:3b}, 
the controller selects \textsc{continue} when both smoothed signals lie 
below threshold, \textsc{update} when a moderate crossing triggers 
lightweight fine-tuning of $f_{\boldsymbol{\theta}}$ after $T_{\text{hold}}$ 
consecutive drift slots (a hold-off count preventing premature 
retraining on transient fluctuations), and \textsc{retire} when a severe 
crossing activates the generative pipeline of 
Section~\ref{subsec:gen_pipeline}. This graded response distinguishes the 
CDT from~\cite{clce_kong2025}, which treats all drift as requiring a new 
basis model. Algorithm~\ref{alg:cdt} summarizes the complete operation.
\begin{algorithm}[t]
\caption{CDT Lifecycle Management}
\label{alg:cdt}
\KwIn{$\mathbf{y}_p^{(t)}$, $f_{\boldsymbol{\theta}}$, $\mathcal{V}_{\boldsymbol{\phi}}$, ref. stats, $\tau_{\rm VAE}$, $\tau_\delta$, $T_{\rm hold}$, $\mathcal{R}$, $\gamma$}
\KwOut{$\hat{\mathbf{H}}^{(t)}$, $\mathcal{A}^{(t)}$}
\textit{Offline}: train $\mathcal{V}_{\boldsymbol{\phi}}$ \eqref{eq:vae_loss}; initiate ref.\ stats \eqref{eq:ref_stats}, thresholds \eqref{eq:thresholds}, $c\leftarrow0$\;
\For{each $t$}{
Estimate $\hat{\mathbf{H}}^{(t)}$, latents via $f_{\boldsymbol{\theta}}$; compute $\bar{\epsilon}_{\rm VAE}(t),\bar{\delta}(t)$ \eqref{eq:vae_error}--\eqref{eq:ewma_delta}\;
\uIf{$\bar{\epsilon}_{\rm VAE}<\tau_{\rm VAE}$ \textbf{and} $\bar{\delta}<\tau_\delta$}{
$\mathcal{A}^{(t)}\leftarrow\textsc{Continue}$, $c\leftarrow0$\;}
\uElseIf{$\bar{\epsilon}_{\rm VAE}<2\tau_{\rm VAE}$ \textbf{and} $\bar{\delta}<2\tau_\delta$}{
$c\leftarrow c+1$; \textbf{if} $c\ge T_{\rm hold}$: fine-tune $f_{\boldsymbol{\theta}}$, $\mathcal{A}^{(t)}\leftarrow\textsc{Update}$, $c\leftarrow0$\;}
\Else{
$\mathcal{A}^{(t)}\leftarrow\textsc{Retire}$, $\hat{\mathbf{H}}^{(t)}\leftarrow\hat{\mathbf{H}}^{\rm LS}_p$\;
\Repeat{${\rm NMSE}'(\mathcal{D}_{\rm syn})\le\gamma$}{
\lIf{$\mathcal{D}_{\rm obs}\in\mathcal{R}$}{load $f_{\boldsymbol{\theta}'}$ from $\mathcal{R}$}
\lElse{generate $\mathcal{D}_{\rm syn}$ (VQ-VAE), train $f_{\boldsymbol{\theta}'}$}
extend $\mathcal{D}_{\rm obs}$ if retrying\;
}
deploy $f_{\boldsymbol{\theta}'}$, update $\mathcal{R}$, reinitiate\ \eqref{eq:ref_stats}, \eqref{eq:thresholds}\;
}
}
\Return{$\hat{\mathbf{H}}^{(t)},\mathcal{A}^{(t)}$}
\end{algorithm}

\subsection{GAI-Driven Synthetic Channel 
Pipeline}
\label{subsec:gen_pipeline}

When \textsc{retire} is triggered, the CDT 
initiates a replacement procedure aligned with 
the GDTC framework in ~\cite{gdtc_liu2025}, adapted here 
for model replacement rather than communication 
parameter selection, and drawing on the 
established role of GAI in compensating for 
data scarcity in DT 
architectures~\cite{chen2024gai_hdt}. A 
VQ-VAE is first trained 
on post-drift pilot observations, 
$\mathcal{D}_{\text{obs}}$, to characterise the 
new channel distribution, with VQ-VAE preferred 
over a standard VAE since its discrete codebook 
naturally captures the sparse multipath 
structure of wireless 
channels~\cite{gdtc_liu2025}. The decoder then 
generates a synthetic dataset, 
$\mathcal{D}_{\text{syn}}$, compensating for 
post-drift data scarcity, on which a replacement 
estimator $f_{\boldsymbol{\theta}'}$ is trained 
offline while the LS fallback maintains 
uninterrupted service. If the new environment 
matches a previously stored scenario, 
$f_{\boldsymbol{\theta}'}$ is retrieved directly 
from the model repository 
$\mathcal{R}$~\cite{khaldi2026agentic}. 
The candidate model is then validated on 
held-out synthetic channels against threshold 
$\gamma$ following the verify-before-deploy 
principle of~\cite{khaldi2026agentic}, and 
deployed only upon passing, with monitoring 
re-initialized thereafter.

\section{Simulation Results}

\begin{table}[!t]
\caption{Simulation Parameters}
\label{tab:sim_params}
\centering
\footnotesize
\setlength{\tabcolsep}{4pt}
\renewcommand{\arraystretch}{0.95}
\begin{tabular}{|p{3.0cm}|p{4.7cm}|}
\hline
\textbf{Parameter} & \textbf{Value} \\
\hline
Channel models (Doppler) & EPA (5 Hz), EVA (70 Hz), ETU (300 Hz) \\
\hline
Carrier / subcarrier spacing / sampling & 2.1 GHz / 15 kHz / 1080 kHz \\
\hline
OFDM numerology & $N_f{=}72$, $N_s{=}14$, CP = 16 samples \\
\hline
Modulation / pilots & QPSK; $N_{pf}{=}3$, $N_p{=}24$ \\
\hline
SNR range & $-5$ to $30$ dB \\
\hline
Estimator & 4-layer FCN (ReLU) \\
\hline
Training data / SNR & 32000 EPA realizations, 12 dB \\
\hline
Test data & 4000 realizations per SNR \\
\hline
VAE training / reference & EPA pilots, 2000 samples \\
\hline
Monitoring ($W$, $\alpha$, threshold) & 100 obs., 0.05, $\bar{\mu}^{\rm cal}{+}2\sigma^{\rm cal}$ \\
\hline
\end{tabular}
\end{table}

We evaluate the proposed framework using the 3GPP channel models of TS~36.101, with OFDM parameters following \cite{attrnet,sccenet} and
simulation parameters listed in Table~\ref{tab:sim_params}. The framework is compared against four baselines: LS, LMMSE (with perfect channel knowledge),
AttRNet~\cite{attrnet}, and SCCENet~\cite{sccenet}. Under matched conditions, LMMSE uses the training-channel statistics, while under EVA/ETU deployment it is given the true deployment-channel statistics, acting as an oracle bound.

\subsection{Estimation Performance Under Matched Conditions}

\begin{figure}[t!]
     \centering
     \begin{subfigure}[b]{0.25\textwidth}
         \centering
         \includegraphics[width=\textwidth]{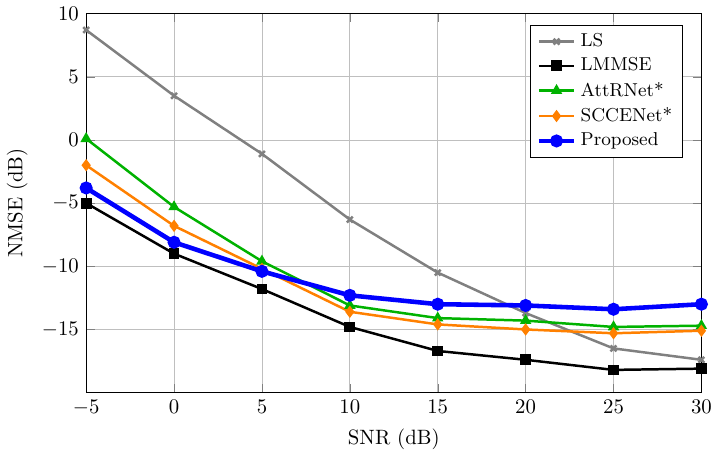}
         \caption{Proposed method trained with lightweight 4-layer FCN.}
         \label{fig:4a}
     \end{subfigure}
     \hfill
     \begin{subfigure}[b]{0.25\textwidth}
         \centering
         \includegraphics[width=\textwidth]{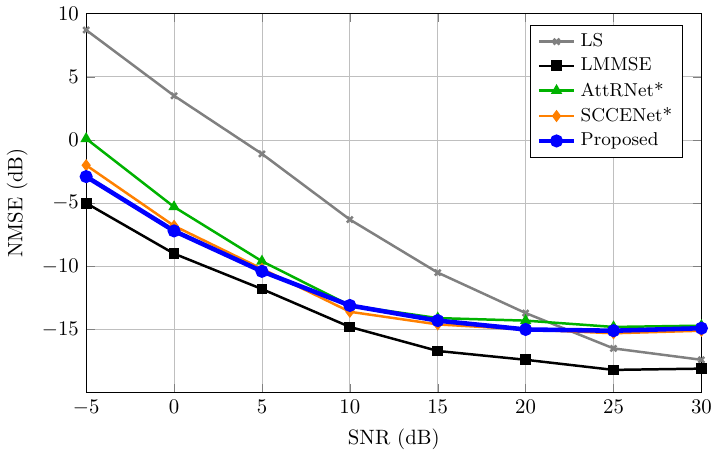}
         \caption{Proposed method trained with SCCENet.}
         \label{fig:4b}
     \end{subfigure}
        \caption{NMSE vs.\ SNR under matched conditions.}
        \label{fig:matched}
\end{figure}

Fig.~\ref{fig:4a} compares the proposed framework using a lightweight FCN backbone under matched EPA training and testing conditions. Although AttRNet and SCCENet achieve slightly lower NMSE at high SNR due to their larger network capacity, the proposed framework maintains competitive performance while using a significantly simpler estimator. Since no channel drift is present, the VAE-based monitoring correctly keeps the lifecycle controller in the \textsc{Continue} state throughout deployment. Fig.~\ref{fig:4b} replaces the FCN with SCCENet while retaining the proposed monitoring and lifecycle management framework. The resulting performance closely matches the original SCCENet, demonstrating that the proposed VAE-based adaptation framework is model-agnostic and can be readily integrated with different deep learning channel estimators, allowing a trade-off between estimation accuracy and computational complexity.

\subsection{Performance Under Distribution Shift}
\begin{figure}[t!]
     \centering
     \begin{subfigure}[b]{0.25\textwidth}
         \centering
         \includegraphics[width=\textwidth]{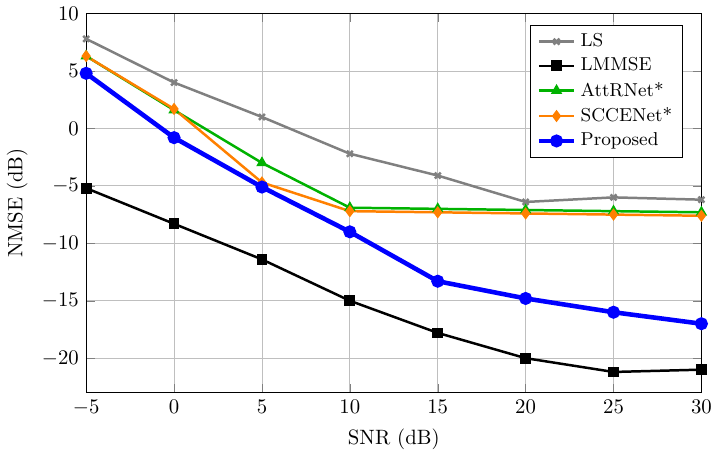}
         \caption{EVA test channel (moderate drift).}
         \label{fig:5a}
     \end{subfigure}
     \hfill
     \begin{subfigure}[b]{0.25\textwidth}
         \centering
         \includegraphics[width=\textwidth]{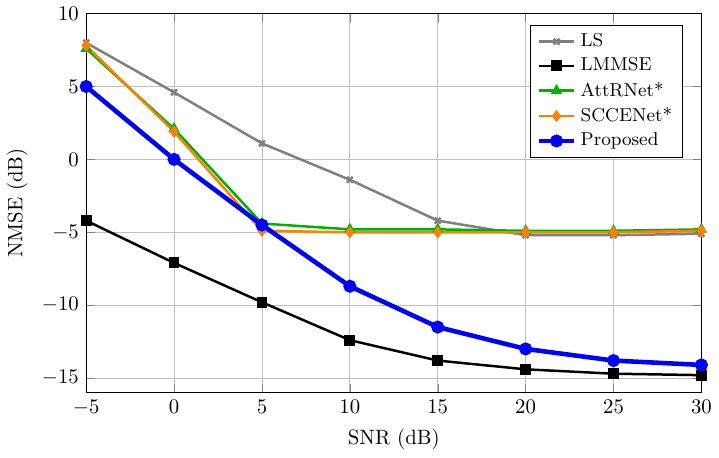}
         \caption{ETU test channel (severe drift).}
         \label{fig:5b}
     \end{subfigure}
        \caption{NMSE vs.\ SNR under distribution shift.}
        \label{fig:drift}
\end{figure}

Fig.~\ref{fig:drift} demonstrates the practical benefit of the proposed
framework, where all learning-based estimators are trained on EPA and
deployed under channel drift without retraining, while LMMSE is assumed to
have perfect knowledge of the deployment channel statistics. Under moderate
EVA drift, AttRNet and SCCENet saturate near $-7$~dB beyond $10$~dB SNR,
unable to adapt to the new statistics. The proposed framework instead
detects the shift via VAE reconstruction error and activation drift,
triggers \textsc{Update}, and continues improving with SNR, reaching
$-17.0$~dB at $30$~dB, only $4$~dB from the oracle LMMSE bound. Under the
more severe ETU channel, AttRNet and SCCENet plateau near $-5$~dB with
little gain over LS, whereas the monitoring module identifies severe drift
and initiates \textsc{Retire}. The VQ-VAE-assisted replacement pipeline then
restores accuracy to $-14.1$~dB at $30$~dB, only $0.8$~dB above LMMSE. These
results confirm that the graded \textsc{Update}/\textsc{Retire} strategy
distinguishes moderate from severe drift, enabling autonomous lifecycle
management with robust estimation performance.

\subsection{Computational Complexity}

The computational complexity of the proposed framework is established by the underlying AI/ML channel estimator.  It increases with the number of network layers and neurons, i.e., $\mathcal{O}(N_{\rm layers}N_{\rm neurons})$ per forward/backward pass. The VAE-based monitoring module introduces only a lightweight auxiliary network and simple threshold evaluation which makes its online inference overhead negligible compared to the backbone estimator. 

\section{Conclusion}
\label{sec:conclusion}

This letter proposed a model-agnostic CDT framework for autonomous lifecycle
management of AI/ML-based channel estimators. By jointly exploiting VAE
reconstruction error and latent activation drift, the framework detects
channel distribution shifts without ground-truth channel information,
achieving robust drift detection and improved estimation under moderate and
severe drift. Future work will extend the framework to massive MIMO systems
and online model replacement via generative digital twins.

\bibliographystyle{IEEEtran}
\bibliography{references}

\end{document}